\documentclass[preprint]{elsarticle}
\makeatletter
\def\ps@pprintTitle{%
 \let\@oddhead\@empty
 \let\@evenhead\@empty
 \def\@oddfoot{}%
 \let\@evenfoot\@oddfoot}
\makeatother
\usepackage{geometry}
\usepackage{tabulary}
\usepackage{fancyhdr}
\fancypagestyle{pprintTitle}{%
\lhead{} \chead{}\rhead{IFIC/15-15, LAL 15-111}
\lfoot{}\cfoot{}\rfoot{}

}
\usepackage[latin1]{inputenc}
\usepackage{graphicx,epsfig}
\usepackage{wrapfig,rotating}
\usepackage{amssymb,amsmath,array}
\usepackage{latexsym, amsfonts, wasysym}
\usepackage{epstopdf}
\usepackage{lscape}
\usepackage{url,color}
\definecolor{orange}{rgb}{1,0.5,0}
\definecolor{gray}{rgb}{0.5,0.5,0.5}
\definecolor{roetlich}{rgb}{1, .7, .7}
\definecolor{camel}{rgb}{0.76, 0.6, 0.42}
\definecolor{britishracinggreen}{rgb}{0.0, 0.26, 0.15}
\usepackage{caption}
\usepackage{subcaption}
\usepackage{appendix}
\usepackage[colorlinks]{hyperref}
\AtBeginDocument{%
  \hypersetup{%
    linkcolor=red,%
    citecolor=green,%
    urlcolor=cyan,%
  }%
}%
\usepackage{feynmf}
\usepackage[]{lineno}

\usepackage[above,section]{placeins}

\usepackage[english]{babel}
\usepackage[T1]{fontenc}

\begin{document}
\newcommand\pubdate{\today}

\def\draftnote#1{{\bf [#1]}}
\newcommand{\bspace}{\!\!\!\!}
\def\met{\mbox{$E{\bspace}/_{T}$}}


\def\stacksymbols #1#2#3#4{\def\theguybelow{#2}
    \def\vp{\lower#3pt}
    \def\sp{\baselineskip0pt\lineskip#4pt}
    \mathrel{\mathpalette\intermediary#1}}

\def\intermediary#1#2{\vp\vbox{\sp
     \everycr={}\tabskip0pt
     \halign{$\mathsurround0pt#1\hfil##\hfil$\crcr#2\crcr
              \theguybelow\crcr}}}


\def\beq{\begin{equation}}
\def\eeq#1{\label{#1}\end{equation}}
\def\eeqn{\end{equation}}
\newcommand{\gsim}{\hbox{ \raise3pt\hbox to 0pt{$>$}\raise-3pt\hbox{$\sim$} }}
\newcommand{\lsim}{\hbox{ \raise3pt\hbox to 0pt{$<$}\raise-3pt\hbox{$\sim$} }}
\newcommand{\mathbold}[1]{\mbox{\boldmath $#1$}}


\newenvironment{Eqnarray}%
   {\arraycolsep 0.14em\begin{eqnarray}}{\end{eqnarray}}
\def\beqa{\begin{Eqnarray}}
\def\eeqa#1{\label{#1}\end{Eqnarray}}
\def\eeqan{\end{Eqnarray}}
\def\CR{\nonumber \\ }


\def\leqn#1{(\ref{#1})}


\def\invfb{ \mbox{fb}^{-1} }
\def\invpb{ \mbox{pb}^{-1} }
\def\roots{ \sqrt{s} }
\def\TeV{ \mbox{TeV}}
\def\GeV{ \mbox{GeV}}
\def\MeV{ \mbox{MeV}}

\def\TeV{\ifmmode {\mathrm{\ Te\kern -0.1em V}}\else
                   \textrm{Te\kern -0.1em V}\fi}%
\def\GeV{\ifmmode {\mathrm{\ Ge\kern -0.1em V}}\else
                   \textrm{Ge\kern -0.1em V}\fi}%
\def\MeV{\ifmmode {\mathrm{\ Me\kern -0.1em V}}\else
                   \textrm{Me\kern -0.1em V}\fi}%
\def\keV{\ifmmode {\mathrm{\ ke\kern -0.1em V}}\else
                   \textrm{ke\kern -0.1em V}\fi}%
\def\eV{\ifmmode  {\mathrm{\ e\kern -0.1em V}}\else
                   \textrm{e\kern -0.1em V}\fi}%
\let\tev=\TeV
\let\gev=\GeV
\let\mev=\MeV
\let\kev=\keV
\let\ev=\eV

\newcommand{\ra}            {\ensuremath{ \rightarrow     }}
\newcommand{\Ra}            {\ensuremath{ \Rightarrow     }}
\newcommand{\longra}        {\ensuremath{ \longrightarrow }}
\newcommand{\Longra}        {\ensuremath{ \Longrightarrow }}
\newcommand{\la}            {\ensuremath{ \leftarrow      }}
\newcommand{\La}            {\ensuremath{ \Leftarrow      }}
\newcommand{\longla}        {\ensuremath{ \longleftarrow  }}
\newcommand{\Longla}        {\ensuremath{ \Longleftarrow  }}
\newcommand{\lra}           {\ensuremath{ \leftrightarrow }}
\newcommand{\Lra}           {\ensuremath{ \Leftrightarrow }}
\newcommand{\longlra}       {\ensuremath{ \longleftrightarrow }}
\newcommand{\Longlra}       {\ensuremath{ \Longleftrightarrow }}
\providecommand{\swsqeffl}    {\sin^2\!\theta_{\rm{eff}}^\ell}
\providecommand{\sstw}{\mbox{$\sin^2\theta_W$}}
\providecommand{\MW}      {m_{\mathrm{W}}}
\providecommand{\MZ}      {m_{\mathrm{Z}}}
\providecommand{\MH}      {m_{\mathrm{H}}}
\providecommand{\Mh}      {m_{\mathrm{h}}}
\providecommand{\MT}      {m_{\mathrm{t}}}
\providecommand{\GZ}      {\Gamma_{\mathrm{Z}}}
\providecommand{\GF}         {G_{\mathrm{F}}}
\providecommand{\ppl}  {{\cal P}_{\rm{e}^+}}
\providecommand{\pmi}  {{\cal P}_{\rm{e}^-}}
\providecommand{\ppm}  {{\cal P}_{\rm{e}^\pm}}
\providecommand{\peff}  {{\cal P}_{\rm{eff}}}
\providecommand{\pol}  {{\cal P}}
\providecommand{\Rb}   {R_{\rm{b}}}
\providecommand{\ee}   {\rm{e^+e^-}}
\providecommand{\bb}   {\rm{b\overline{b}}}
\providecommand{\thw}   {\theta_{W}}
\providecommand{\nb}{\,\mathrm{nb}}
\providecommand{\pb}{\,\mathrm{pb}}
\providecommand{\fb}{\,\mathrm{fb}}\providecommand{\pbi}{\,\mathrm{pb}^{-1}}
\providecommand{\fbi}{\,\mathrm{fb}^{-1}}
\providecommand{\Cdgz}{\ensuremath{\Delta g^\mathrm{Z}_1}}
\providecommand{\Cdgg}{\ensuremath{\Delta g^\mathrm{\gamma}_1}}
\providecommand{\Cdkz}{\ensuremath{\Delta \kappa_\mathrm{Z}}}
\providecommand{\Cdkg}{\ensuremath{\Delta \kappa_{\gamma}}}
\providecommand{\Ckg}{\ensuremath{\kappa_{\gamma}}}
\providecommand{\Ckz}{\ensuremath{\kappa_{\mathrm{Z}}}}
\providecommand{\Clg}{\ensuremath{\lambda_{\gamma}}}
\providecommand{\Clz}{\ensuremath{\lambda_{\mathrm{Z}}}}
\providecommand{\Cgv}[1]{\ensuremath{g^V_{#1}}}
\providecommand{\Cgz}[1]{\ensuremath{g^Z_{#1}}}
\providecommand{\Cgg}[1]{\ensuremath{g^{\gamma}_{#1}}}
\providecommand{\Ckzt}{\ensuremath{\tilde{\kappa}_\mathrm{Z}}}
\providecommand{\Clzt}{\ensuremath{\tilde{\lambda}_\mathrm{Z}}}
\providecommand{\Ckgt}{\ensuremath{\tilde{\kappa}_{\gamma}}}
\providecommand{\Clgt}{\ensuremath{\tilde{\lambda}_{\gamma}}}
\providecommand{\phmi}{\phantom{-}}
\newcommand{\sw}{s_w}
\newcommand{\cw}{c_w}
\newcommand{\ii}{\mathrm{i}}
\newcommand{\vA}{\mathbf{A}}
\newcommand{\vB}{\mathbf{B}}
\newcommand{\vC}{\mathbf{C}}
\newcommand{\vD}{\mathbf{D}}
\newcommand{\vV}{\mathbf{V}}
\newcommand{\vT}{\mathbf{T}}
\newcommand{\vW}{\mathbf{W}}
\newcommand{\vw}{\mathbf{w}}
\newcommand{\cD}{\mathcal{D}}
\newcommand{\cS}{\mathcal{S}}
\newcommand{\cP}{\mathcal{P}}
\newcommand{\pd}{\partial}
\newcommand{\mfrac}[2]{#1 / #2}
\newcommand{\tr}[1]{\mathop{\rm tr}\left\{#1\right\}}
\newcommand{\LL}{\mathcal{L}}
\newcommand{\MSbar}{\mbox{$\overline{\rm MS}$}}
\newcommand{\pp}{{\prime 2}}
\newcommand{\z}{\phantom{0}}

\newcommand{\polRL}{e^-_R e^+_L}
\newcommand{\polLR}{e^-_L e^+_R}
\newcommand{\ttbar}{ t \bar t}
\newcommand{\Gammat}{\Gamma_t}
\newcommand{\rmsn}{\mathrm{rms}_{90}}
\newcommand{\Pt}{P_T}
\newcommand{\eplus}{e^+}
\newcommand{\eminus}{e^-}
\newcommand{\epem}{\eplus\eminus}
\def\invfb{ \mbox{fb}^{-1} } 
\newcommand{\alr}{A_{LR}}
\newcommand{\afb}{A_{FB}}
\newcommand{\afbt}{A^t_{FB}}
\newcommand{\thel}{\theta_{hel}}
\newcommand{\cthel}{\mathrm{cos} \theta_{hel}}
\newcommand{\qq}{q\bar{q}}
\newcommand{\Zzero}{Z^0}
\newcommand{\tpq}{t}
\newcommand{\Wboson}{W}
\newcommand{\bottom}{b}
\newcommand{\quark}{q}
\newcommand{\qll}{Q_{LL}}
\newcommand{\qlr}{Q_{LR}}
\newcommand{\qrl}{Q_{RL}}
\newcommand{\qrr}{Q_{RR}}
\newcommand{\leftp}{\left(}
\newcommand{\rightp}{\right)}
\newcommand{\lhel}{\lambda_t}
\newcommand{\fonevI}{{\cal F}^{I}_{1V}}
\newcommand{\ftwovI}{{\cal F}^{I}_{2V}}
\newcommand{\foneaI}{{\cal F}^{I'}_{1A}}
\newcommand{\ftwoaI}{{\cal F}^{I}_{2A}}
\newcommand{\pem} { {\cal P}_{\eminus} } 
\newcommand{\pep} {{\cal P}_{\eplus} }
\newcommand{\fonevZ}{F^{Z}_{1V}}
\newcommand{\fonevZh}{F^{Z}_{1V}}
\newcommand{\ftwovZ}{F^{Z}_{2V}}
\newcommand{\foneaZ}{F^{Z}_{1A}}
\newcommand{\ftwoaZ}{F^{Z}_{2A}}
\newcommand{\fonevA}{F^{\gamma}_{1V}}
\newcommand{\fonevAh}{F^{\gamma}_{1V}}
\newcommand{\ftwovA}{F^{\gamma}_{2V}}
\newcommand{\foneaA}{F^{\gamma}_{1A}}
\newcommand{\ftwoaA}{F^{\gamma}_{2A}}
\newcommand{\glZ}{g^{Z}_L}
\newcommand{\grZ}{g^{Z}_R}
\newcommand{\glA}{g^{\gamma}_L}
\newcommand{\grA}{g^{\gamma}_R}

\newcommand{\afbl}{(A_{FB})_L}
\newcommand{\afbr}{(A_{FB})_R}
\newcommand{\afbI}{(A_{FB})_I}

\begin{frontmatter}

\title{\LARGE\bf  A precise characterisation of the top quark electro-weak vertices at the ILC}
\author[add1]{M.S.\,Amjad\fnref{fn1}}
\author[add1]{S.\,Bilokin} 
\author[add2]{M.\,Boronat}
\author[add1]{P.\,Doublet\fnref{fn3}}
\author[add1]{T.\,Frisson\fnref{fn2}}
\author[add2]{I.~Garc\'{\i}a{~}Garc\'{\i}a  }
\author[add2]{M.\,Perell\'o}
\author[add1]{R.\,P\"oschl\corref{cor1}}
\author[add1]{F.\,Richard}
\author[add2]{E.\,Ros}
\author[add1]{J.~Rou\"en\'e}
\author[add2]{P.\,Ruiz Femenia\fnref{fn4}}
\author[add2]{M.\,Vos}

\address[add1]{Laboratoire de l'Acc\'el\'erateur Lin\'eaire (LAL), Centre Scientifique d'Orsay, Universit\'e Paris-Sud XI, BP 34, B\^atiment 200, F-91898 Orsay CEDEX, France}
\fntext[fn1]{Now at COMSATS Institute of Information Technology, Islamabad, Pakistan.}
\fntext[fn2]{Now at CERN, 1211 Gen\`{e}ve 23, Switzerland.}
\fntext[fn3]{Now at IUT d'Orsay (Universit\'e Paris-Sud), France.}
\address[add2]{IFIC, Universitat de Valencia CSIC, c/ Catedr\'atico Jos\'e Beltr\'an, 2  46980 Paterna, Spain.}
\fntext[fn4]{Now at Technische Universit\"at M\"unchen, 85748 Garching, Germany.}
\cortext[cor1]{Corresponding author: poeschl@lal.in2p3.fr}



\begin{abstract}
Top quark production in the process $e^+e^- \rightarrow t\bar{t}$ at a future linear electron positron collider with polarised beams is a powerful tool to determine indirectly the scale of new physics.  The presented study, based on a detailed simulation of the ILD detector concept, assumes a centre-of-mass energy of $\roots=500$\,GeV and a luminosity of $\mathcal{L}=500\,\invfb$ equally shared between the incoming beam polarisations of $\pem, \pep =\pm0.8,\mp0.3$. Events are selected in which the top pair decays semi-leptonically and the cross sections and the forward-backward asymmetries are determined.  Based on these results, the vector, axial vector and tensorial $CP$ conserving couplings are extracted separately for the photon and the $Z^0$ component. With the expected precision, a large number of models in which the top quark acts as a messenger to new physics can be distinguished with many standard deviations. This will dramatically improve expectations from e.g. the LHC for electro-weak couplings of the top quark.   
 
\end{abstract}

\end{frontmatter}

\def\thefootnote{\fnsymbol{footnote}}
\setcounter{footnote}{0}
%

\section{Introduction}\label{sec:intro}
The main goal of current and future machines at the energy frontier is to understand the nature of electro-weak symmetry breaking.
This symmetry breaking can be generated by the existence of a new strong sector, inspired by QCD, that may manifest itself at energies of around 1\,TeV. In all realisations of the new strong sector, as for example Randall-Sundrum models~\cite{Randall:1999ee} or compositeness models~\cite{Pomarol:2008bh}, the strength of the coupling to this new sector of the Standard Model fields are supposed to increase with their mass. For this and other reasons, the heavy top quark or $t$~quark with a mass of approximately $m_t =173\GeV$~\cite{Agashe:2014kda} is expected to be a window to any new physics at the TeV energy scale. New physics can modify the electro-weak $\ttbar X$ vertex described in the Standard Model by {\em V}ector and {\em A}xial vector couplings $V$ and $A$ to the vector bosons $X=\gamma, Z^0$.
At the International Linear Collider, ILC~\cite{bib:ilc-tdr-dbd},  that will collide electron and positrons at a centre-of-mass energy of $500\GeV$, $t$~quark electro-weak couplings can be measured at the \% level. 


In contrast to the situation at hadron colliders, the leading-order pair production process $e^+e^- \rightarrow t\bar{t}$ goes directly through the $t\bar{t} Z^0$ and $t\bar{t} \gamma$ vertices.  There is no concurrent QCD production of $t$~quark pairs, which increases greatly the potential for a clean measurement. 
A parametrisation of the $t\bar{t} X$ vertex valid to all orders of perturbation theory may be written as 
\footnote{A dependence on an additional term $(q+\bar{q})_{\mu} \cdot F_3$ can be neglected in the limit of a vanishing electron mass~\cite{Kane:1991bg}.}: 

\begin{equation}
\Gamma^{\ttbar X}_{\mu}(k^2, q, \bar{q}) = ie\left\{\gamma_{\mu}\left( F_{1V}^X(k^2) +\gamma_5 F_{1A}^X(k^2) \right)   - \frac{\sigma_{\mu\nu}}{2m_t}(q+\bar{q})^{\nu}  \left(  iF_{2V}^X(k^2) +\gamma_5 F_{2A}^X(k^2)  \right) \right\},
\label{eq:vtxvtt}
\end{equation}
with $e$ being the electrical charge of the electron, $k^2=(q+\bar{q})^2$ being the squared four-momentum of the exchanged boson and $q$ and $\bar{q}$ being the four-vectors of the $t$ and $\bar{t}$~quark, respectively. Further, $\gamma_\mu$  are the Dirac matrices leading to vector currents of fermions and $\gamma_5$ is the Dirac matrix allowing to introduce an axial-vector current into the theory. Finally, $\sigma_{\mu\nu}=\frac{i}{2}\leftp \gamma_{\mu}\gamma_{\nu} - \gamma_{\nu}\gamma_{\mu} \rightp$ allows for describing the scattering of a particle with spin 1/2 and a given magnetic moment.


Within the Standard Model the $F_1$ have the following values at tree level:
\begin{equation}
F_{1V}^{\gamma,SM}=\frac{2}{3},\,F_{1A}^{\gamma,SM}=0,\,F_{1V}^{Z,SM}=\frac{1}{4s_wc_w}\left(1-\frac{8}{3}s^2_w \right),\,F_{
1A}^{Z,SM}=-\frac{1}{4s_wc_w},
\label{eq:ffactors}
\end{equation}
while all the $F_2$ are zero. In Eq.~\ref{eq:ffactors} $s_w$ and $c_w$ are the sine and the cosine of the Weinberg angle $\theta_W$. The scale dependence of the form factors is a consequence of higher order corrections. The corrections of the vector currents lead to the anomalous electro-magnetic and weak-magnetic  moments represented by $F^{X}_{2V}$ that correct the gyromagnetic ratio $g_t$ of the $t$ quark. Typical values for these corrections are in the range ${\cal O}(10^{-3} - 10^{-2})$~\cite{Labun:2012fg}.  Corrections to the axial-vector current result in the Form Factors $F^{X}_{2A}$ that are related to the dipole moment $d_{t}^{X}=(e/2m_t)F^{X}_{2A}(0)$ that in turn violates the combined {\em C}harge and {\em P}arity symmetry $CP$. Otherwise said, all couplings but $F_{2A}^X(k^2)$ conserve $CP$. 

The Form Factors $F^{Z}_{1V,A}$ are related to couplings of $t$ quarks with left and right-handed helicity to the $Z^0$:

\begin{equation}
g^{Z}_L = \fonevZ - \foneaZ,
\qquad
g^{Z}_R=\fonevZ + \foneaZ
\label{eq:qtrl}
\end{equation}
Trivially, the same equations apply correspondingly to the photon couplings $g^{\gamma}_L$

In this paper the precision of $CP$ conserving form factors and couplings as introduced above will be derived by means of a full simulation study of the reaction $e^+e^- \rightarrow t\bar{t}$ at a centre-of-mass energy of $\roots=500\GeV$ with 80\% polarised electron beams and 30\% polarised positron beams using experimentally well defined observables. Special emphasis will be put on the selection efficiency and the polar angle of the final state $t$ quarks. Both experimental quantities are suited to monitor carefully experimental systematics that may occur in the extraction of form factors and couplings.  
The results presented in the following are based on the studies described in detail in Refs.~\cite{Amjad:2013tlv,bib:these-rouene}.
\section{Top quark production at the ILC}

\begin{figure}[h!]
\begin{subfigure}{0.5\textwidth}
\setlength{\abovecaptionskip}{24pt plus 0pt minus 0pt}
\centering
\begin{fmffile}{fmfeegff}
\begin{fmfchar*}(36,27)
\fmfleft{em,ep} \fmflabel{$\eplus$}{ep} \fmflabel{$\eminus$}{em}
\fmfright{f,fbar} \fmflabel{$t$}{f} \fmflabel{$\bar{t}$}{fbar}
\fmf{fermion}{em,Zee,ep}
\fmf{fermion}{fbar,Zff,f}
\fmf{photon,label=$\gamma^{\ast}/Z^{0 \ast}$}{Zee,Zff}
\fmfdot{Zee,Zff}
\end{fmfchar*}
\end{fmffile}
\caption{$t\bar{t}$ pair production}
\label{fig:diagrams_a}
\end{subfigure}
\begin{subfigure}{.5\textwidth}
\setlength{\abovecaptionskip}{24pt plus 0pt minus 0pt}
\centering
\begin{fmffile}{fmfeest}
\begin{fmfchar*}(36,27)
\fmfleft{i1,i2}  \fmflabel{$\eminus$}{i1} \fmflabel{$\eplus$}{i2}
\fmfright{o1,o2,o3}  \fmflabel{$\bar{b}$}{o2} \fmflabel{$W^-$}{o1} \fmflabel{$t$}{o3} 
\fmf{fermion}{i1,v1}
\fmf{fermion, label=$\nu^{\ast}_e$,l.side=left}{v1,v3}
\fmf{fermion}{v3,i2}
\fmf{photon}{v3,v2} 
\fmfL(23.10751,25.88596,t){$W^{+ \ast }$}
\fmf{fermion}{o2,v2,o3} 
\fmf{photon}{v1,o1} 
\fmfdot{v1,v2,v3}
\fmffreeze
\fmfshift{0.04w,0.1w}{v3}
\fmfshift{0.0w,0.05w}{v2}
\fmfshift{0.0w,0.05w}{o2}
\fmfshift{0.1w,0.0w}{o3}
\end{fmfchar*}
\end{fmffile}
\caption{Single $t$ quark production}
\label{fig:diagrams_c}
\end{subfigure}
\caption{\sl Diagrams that contribute to the $e^+ e^- \rightarrow l \nu b \bar{b} q' \bar{q}$ production: (a) Tree level $t \bar{t}$ pair production, (b) single $t$~quark production. }
\label{fig:feyn_diagrams}
\end{figure}
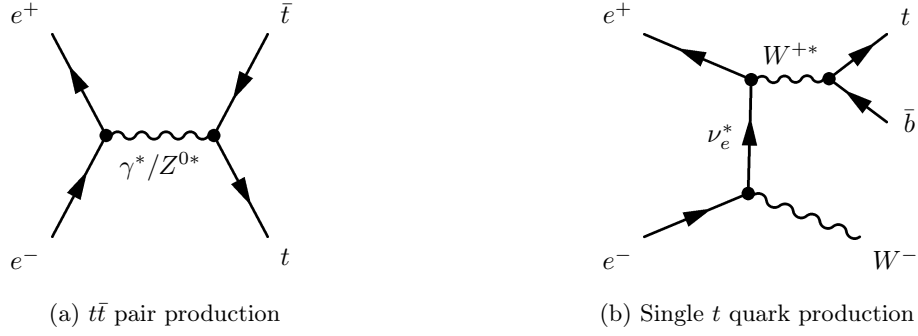

The tree level diagram for pair production of $t$ quarks at the ILC is presented in Figure~\ref{fig:diagrams_a}. 
The decay of the top quarks proceeds predominantly through  $t \rightarrow W^{\pm} b$. The subsequent decays of the $W^{\pm}$ bosons to a charged lepton and a neutrino or a quark-anti-quark pair lead to a  six-fermion final state. The study presented in this article focuses on the 'lepton+jets' final state $ l^{\pm} \nu b \bar{b} q' \bar{q}$ representing a branching fraction of about 43.4\% on all $t\bar{t}$~pair decays. 


\setlength{\unitlength}{1.0mm}

Several other Standard Model processes give rise to the same final state. The most important source is single $t$~quark production through the process
$e^+e^- \rightarrow W W^* \rightarrow W t \bar{b} \rightarrow l^{\pm} \nu b \bar{b} q' \bar{q}$. One of the diagrams contributing to this process is
presented in Figure~\ref{fig:diagrams_c}. Another relevant source is the $Z^0 W^+ W^-$ production. Due to the coupling of initial state electrons or positrons to $W$ bosons both sources contribute nearly exclusively in a configuration with left-handed polarised electron beams and right-handed polarised positron beams. 

In that case single $t$~quark and $Z^0 W^+ W^-$~boson production can yield a total production rate of up to 10\% of that of the pair production diagram of Fig.~\ref{fig:diagrams_a}. Experimentally, $Z^0 W^+ W^-$ production can be distinguished rather efficiently from $t\bar{t}$ pair production, but a clean separation of final states with a single $t$ quark seems impossible. A realistic experimental strategy must therefore consider the $W^+ b W^- \bar{b}$ inclusively~\cite{Fuster:2015ly}.

\subsection{Observables and form factors}

In case of polarised beams Ref.~\cite{Schmidt:1995mr} suggests to express the form factors introduced in Sec.~\ref{sec:intro} in terms of the helicity of the incoming electrons,  
\begin{eqnarray}
{\cal F}^L_{ij}&=& -F^\gamma_{ij}+
\Bigl({-{1\over2}+s_w^2\over s_wc_w}\Bigr)\Bigl({s\over
s-m_Z^2}\Bigr)F^Z_{ij}\nonumber\\
{\cal F}^R_{ij} &=&-F^\gamma_{ij}+\Bigl({s_w^2\over s_wc_w}\Bigr)
\Bigl({s\over s-m_Z^2}\Bigr)
F^Z_{ij}\ ,\label{combffs}
\end{eqnarray}
with $i=1,2$ and $j=V,A$ and $m_Z$ being the mass of the $Z^0$~boson.
The tree level cross section for $\ttbar$~quark pair production for an electron beam polarisation $I=L,R$ reads 
\beq
\sigma_{I}=2{\cal A}N_c\beta\left[  (1+0.5\gamma^{-2})(\fonevI)^2  + (\foneaI)^2+3\fonevI \ftwovI   + (1+0.5\gamma^{2}) (\ftwovI)^2  \right],
\label{eq:sigma}
\eeqn
where ${\cal A}=\frac{4\pi\alpha^2}{3s}$ with the running electromagnetic coupling $\alpha(s)$ and $N_c$ is the number of quark colours. Furthermore $\gamma$ and $\beta$ are the Lorentz factor and the velocity of the $t$~quark, respectively. The term $\foneaI = \beta {\cal F}^{I}_{1A}$ describes the reduced sensitivity to axial vector couplings near the $\ttbar$ production threshold. The cross sections at the Born level of the signal process $\epem \rightarrow \ttbar$ and the main Standard Model background processes at a centre-of-mass energy of $500\GeV$ are summarised in Table~\ref{tab:procs}. 
\begin{table}[!h]
\begin{center}
\centerline{\begin{tabular}{|c|c|c|c|}
\hline
Channel & $\sigma_{unpol.}$ [fb]& $\sigma_{-,+}$ [fb]& $\sigma_{+,-}$ [fb] \\ 
\hline
$\ttbar$ & 572 & 1564 & 724 \\
\hline
$\mu^+ \mu^-$ & 456 & 969 & 854 \\
\hline
$\sum_{\mathrm{q=u,d,s,c}} \qq$ & 2208 & 6032 & 2793 \\
\hline
$b \bar{b}$ & 372 & 1212 & 276 \\
\hline
$\gamma \Zzero$ & 11185 & 25500 & 19126 \\
\hline
$W^+ W^-$ & 6603 & 26000 & 150 \\
\hline
$\Zzero \Zzero$ & 422 & 1106 & 582 \\
\hline 
$\Zzero W^+ W^-$ & 40 & 151 & 8.7\\
\hline
$\Zzero \Zzero \Zzero$ & 1.1 & 3.2 & 1.22 \\
\hline
Single $t$ for $e^+e^- \rightarrow e^- \bar{\nu_e} t \bar{b} $~\cite{Boos:2001sj} & 3.1  & 10.0 & 1.7\\ 
\hline
\end{tabular}}
\end{center}
\caption{\sl Unpolarised cross-sections and cross-sections at tree level for 100\% beam polarisation for signal and background processes.}
\label{tab:procs}
\end{table}%

The forward-backward asymmetry $\afbt$ can be expressed as 
\beq
(\afbt)_{I} =\\ \mp {\cal A}N_c\beta \cdot  \frac{3 \foneaI (\fonevI + \ftwovI )}{ \sigma_I  }.
\label{eq:afbt}
\eeqn
The '-' sign applies in case of an initial left-handed polarised electron beam, i.e. $I=L$, and the '+' applies correspondingly in case of an initial right-handed polarised electron beam, i.e. $I=R$. In the Standard Model the forward-backward asymmetry takes the values $(\afbt)_{L} =0.37 $ and  $(\afbt)_{R} =0.45$ at tree level.

Neglecting $CP$ violating form factors, the fraction of right-handed $t$ quarks is given by the following expression:

\beq
(F_R)_{I} = 0.5 \mp \frac{2}{3} (\afbt)_{I} 
\label{eq:fri}
\eeqn
The '-' sign applies in case of an initial left-handed polarised electron beam, i.e. $I=L$, and the '+' applies correspondingly in case of an initial right-handed polarised electron beam, i.e. $I=R$. The values expected in the Standard Model at tree level are $(F_R)_{L}=0.25$ and $(F_R)_{R}=0.80$. 

With the introduced observables the six $CP$ conserving form factors defined for the $Z^0$ and the photon can in principle be extracted simultaneously. However, close to the $\ttbar$ threshold the observables depend always on the sum $F_{1V}+F_{2V}$. Therefore, a full disentangling of the form factors will be imprecise for energies below about $1\TeV$. Hence, in the present study either the precision on the Form Factors $F^{X}_{1V,A}$, or equivalently on the Couplings $g^{X}_{L,R}$, are determined simultaneously, while the two $F_{2V}$ are kept at their Standard Model values or vice versa. Due to these considerations the study will only make use of the cross section and $\afbt$ since these are either the most precise observable in case of the cross section or the one that is most sensitive to axial couplings in case of $\afbt$. 
It is however reminded that in~\cite{Amjad:2013tlv} the fraction of right-handed $t$ quarks is determined to a precision of about 2\%.

\subsection{Theory uncertainties}
\label{sec:theory}

The extraction of form factors requires precise predictions of the 
inclusive top quark pair production rate and of several differential 
distributions. In this section the state-of-the-art
calculations and estimate theoretical uncertainties are briefly reviewed.

As discussed at the beginning of this section, the optimal experimental strategy should consider
$e^+e^- \rightarrow W^+ bW^- \bar{b}$ inclusively, without attempting to distinguish single top and top quark
pair production. However, today, sufficiently precise calculations are not available for
the full process $e^+e^- \rightarrow  W^+ bW^- \bar{b}$. Therefore the discussion in this paper is based  on the current
state of the art calculations for $e^+e^- \rightarrow t \bar{t}$, assuming that in the next decade theorists
will rise to the challenge of extending the calculations to $e^+e^- \rightarrow W^+ bW^- \bar{b}$.

The QCD corrections to $e^+e^- \rightarrow t \bar{t}$ production are
known up to $N^3LO$ for the inclusive cross section~\cite{Kiyo:2009gb}, 
and to $NNLO$ for the forward backward asymmetry 
$\afbt$~\cite{Bernreuther:2006vp}. The perturbative series shows good convergence. 

In the kinematic region at around $\roots = 500\GeV$ as relevant for this study the $N^3LO$ correction to the total cross-section is below 1 \%. An estimate 
of the size of the next order - obtained from the conventional variation of the renormalisation scale by a factor two and one half - yields 0.3 \%. It can therefore be
concluded that the uncertainty of today's state-of-the-art calculations is at the per mil level.

In a similar manner the QCD corrections to the prediction of differential distributions and quantities such as the forward-backward 
asymmetry can be estimated. The size of the $N^3LO$ correction to $\afbt$ is estimated using the scale variation to be smaller than 1\%, see also the discussion in e.g.~\cite{Bernreuther:2006vp}.

Electro-weak (EW) corrections to the same process have also been calculated. 
A full one-loop calculation is presented in~\cite{Fleischer:2003kk}.
The correction to the total cross section is found to be approximately 5\%.
The electro-weak correction to the forward-backward asymmetry is large, approximately
10\%~\cite{Fleischer:2003kk,Khiem:2012bp}. Recent studies~\cite{Khiem:2015ofa} show that the corrections are notably different for different
beam polarisations. They change for example the shape of the angular distribution in case of $\pem, \pep =-1, +1$ beam polarisation while they only 
influence the normalisation in case of  $\pem, \pep =+1, -1$. 


The above discussion refers to corrections to the 
process $e^+e^- \rightarrow t \bar{t}$. Further corrections 
of order $\Gamma_t/m_t \sim $ 1\% are expected to appear if the decay of the top
quarks is included in the calculation.

It can be concluded that the state-of-the-art calculations of QCD corrections offer 
the precision required for this study. Uncertainties
are under relatively good control, with uncertainties to the cross section
of the order of a few per mil and order 1\% on the forward-backward asymmetry. Electro-weak (one-loop) corrections are large.
Further work is needed to estimate the size of the two-loop correction 
and, ultimately, to calculate this contribution. Currently these aspects are discussed with theory groups.

\section{Analysis of simulated events}\label{sec:exp}



The study has been carried out on a fully polarised sample albeit realistic values for the ILC are $\pem, \pep =\pm0.8,\mp0.3$. 
The cross section and therefore its uncertainty scales with the polarisation in a well defined way according to~\cite{MoortgatPick:2005cw} 

\beq
\sigma_{\pem,\pep} = \frac{1}{4}\left[(1- \pem \pep)(\sigma_{-,+}+\sigma_{+,-})+(\pem-\pep)(\sigma_{+,-}-\sigma_{-,+})\right]. 
\label{eq:tot-cross}
\eeqn

The observable $\afbt$ varies only very mildly with the beam polarisation. The realistic beam polarisation will be correctly taken into account in the uncertainty of the results. 

Signal and background events corresponding to a luminosity of $250\,\invfb$ at $\roots = 500\GeV$ for each of the two polarisation configurations are generated with version 1.95 of the {\tt WHIZARD} event  generator~\cite{Kilian:2007gr,Moretti:2001zz} that provides lowest order calculations of the $2\rightarrow 6$ fermions subprocess and simulates multiple photon radiation from the initial state electron and positron in leading-logarithmic approximation. {\tt WHIZARD} is interfaced to the {\tt PYTHIA} Monte Carlo programme~\cite{Sjostrand:2006za} for QCD and QED showering. The generated events were subject to a full simulation of the ILD Detector~\cite{bib:ilc-tdr-dbd} and subsequent event reconstruction using the version {\tt ILD\_o1\_v05} of the ILD software.  



The decay of the $t$~quarks proceeds predominantly through  $t \rightarrow W^{\pm} b$. The subsequent decays of the $W^{\pm}$ bosons to a charged lepton and a neutrino or a quark-anti-quark pair lead to a  six-fermion final state. The study presented in this article focuses on the semi-leptonic final state $ l^{\pm} \nu b \bar{b} q' \bar{q}$.  Several other Standard Model processes give rise to the same final state.
The most important source is single $t$~quark production. Another relevant source is the $Z^0 W^+ W^-$ production.
Experimentally,  $Z^0 W^+ W^-$ production can be distinguished rather efficiently from $t$~quark pair production. The separation between single $t$~quark production and $t\bar{t}$ pair production is much more involved. Note however, that according to Table~\ref{tab:procs} single $t$~quark production is strongly suppressed in case of $\pem, \pep =+1,-1$.

The entire selection procedure including lepton and $b$ jet identification, top quark reconstruction and suppression multi-peripheral $\gamma \gamma \rightarrow ${\it hadrons} background is explained in detail in~\cite{Amjad:2013tlv,bib:these-rouene} and~\cite{Boronat:2014hva}. The total selection efficiency of about 56\% for semi-leptonic $\ttbar$ events includes events with a $\tau$~lepton in the final state.  Background processes can be very efficiently removed down to a negligible level. A powerful tool is the $b$~likeness or $b$-tag value that suppresses about 97\% of the dominant $W^+  W^-$ background. Additional selection criteria comprise cuts on the $t$~quark and $W^{\pm}$~boson masses and of the invariant mass of the total hadronic final state. 

With the determined efficiencies a statistical uncertainty of the cross section $e^+e^- \rightarrow t\bar{t}$  of 0.47\% in case $\pem, \pep = -0.8, +0.3$  and 0.63\% in case $\pem, \pep = +0.8, -0.3$ can be derived. 

\subsection{Forward-Backward Asymmetry $\afbt$}\label{sec:extrac}

\begin{figure}[tbp]
\begin{center}
\includegraphics[width=0.49\textwidth]{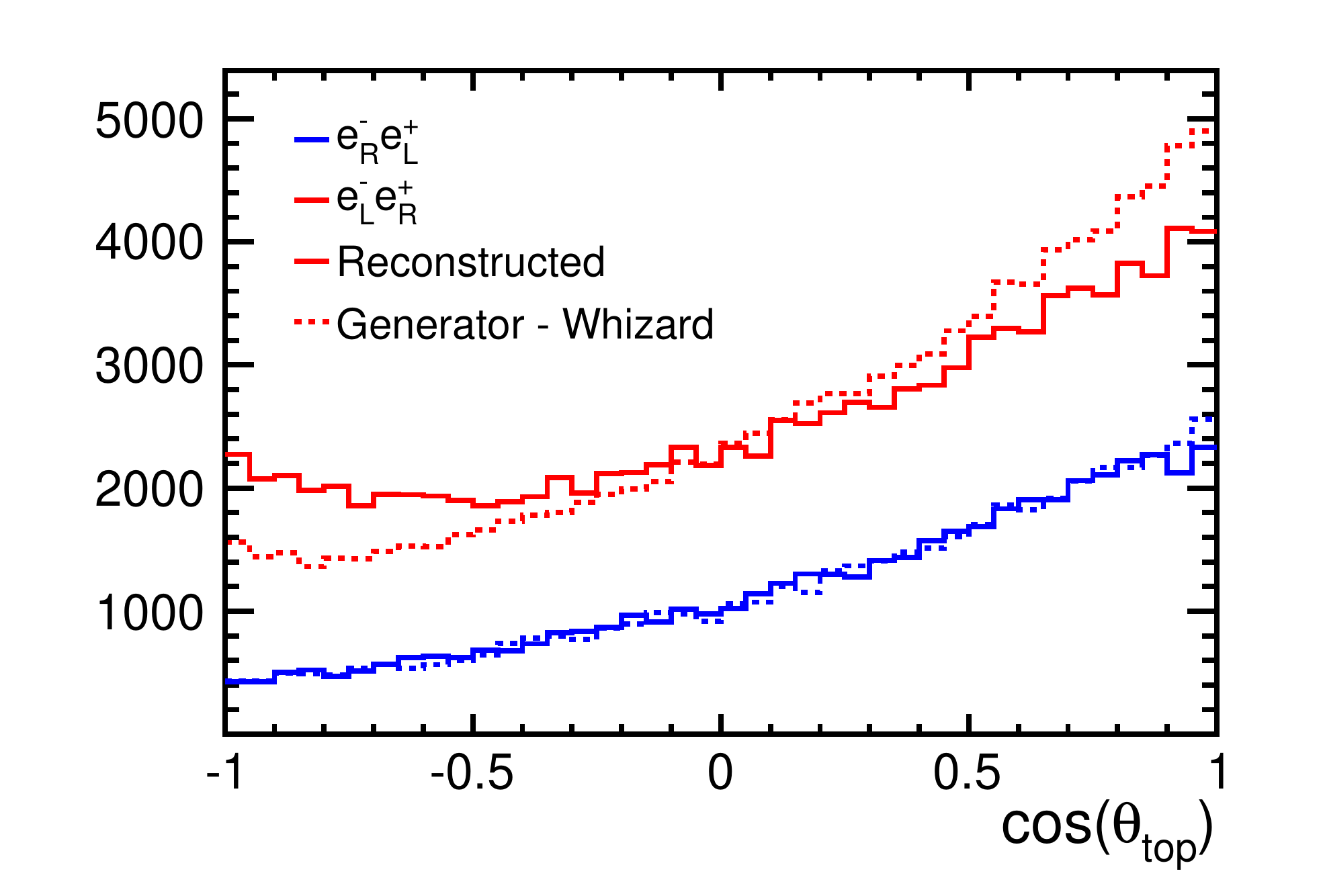}
\includegraphics[width=0.49\textwidth]{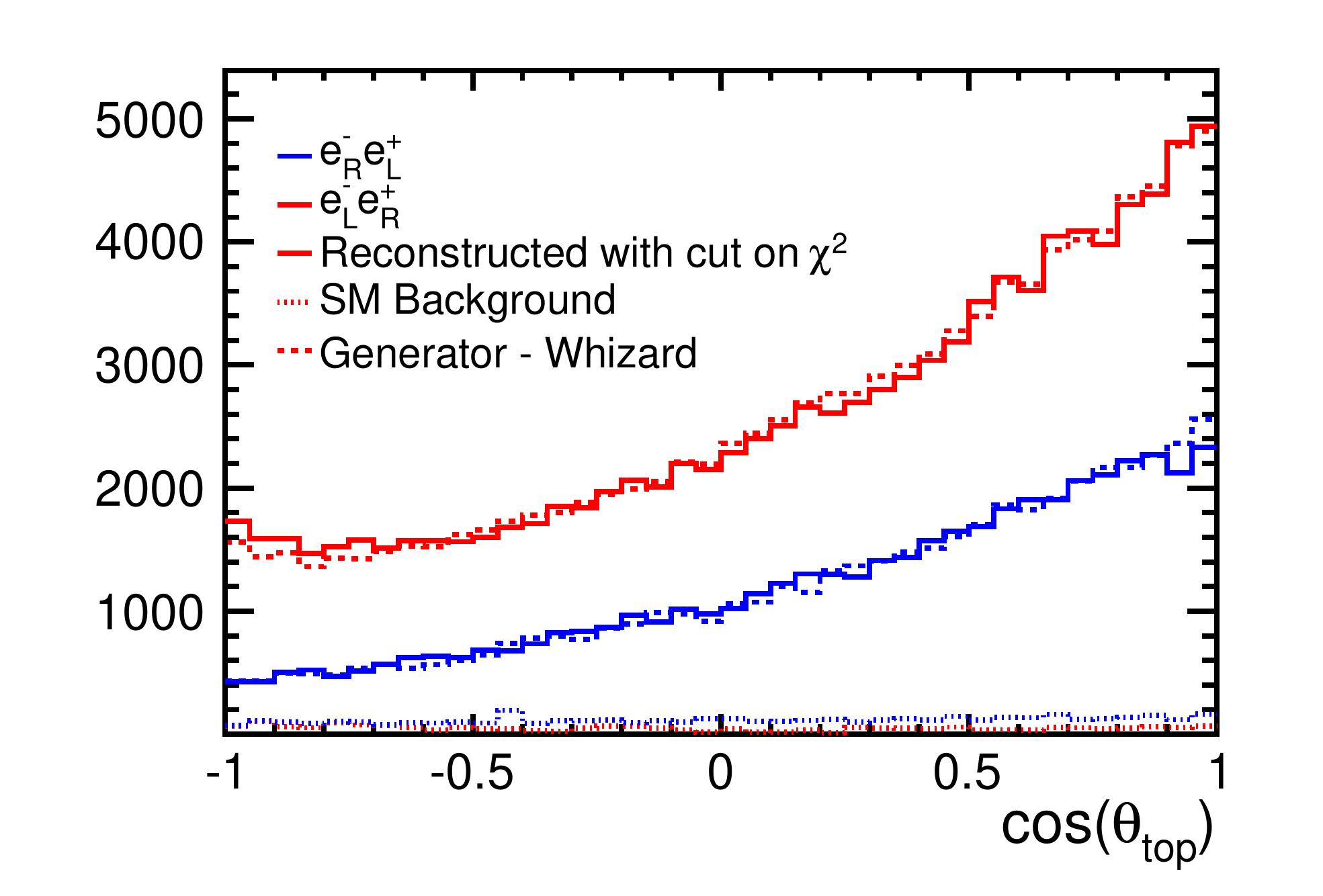}
\caption{\sl  \underline{Left:} Reconstructed forward-backward asymmetry compared with the prediction by the event generator {\sc WHIZARD} {\protect \cite{Kilian:2007gr}} for two configurations of the beam polarisations. \underline{Right:} The same but after the application of a on $\chi^2<15$ for the beam polarisations $\pem,\pep =-1,+1$ as explained in the text. Note, that in both figures no correction is applied for the beam polarisations $\pem,\pep =+1,-1$. The figure on the right hand side shows also the residual Standard Model background.}
\label{fig:ambig_rec}
\end{center}
\end{figure}

The forward-backward asymmetry $\afbt$ has the well known definition 
\beq
\afbt = \frac{N(\mathrm{cos}\theta_{top}>0)-N(\mathrm{cos}\theta_{top}<0)}{N(\mathrm{cos}\theta_{top}>0)+N(\mathrm{cos}\theta_{top}<0)},
\eeqn
where $N$ is the number of events in the two detector hemispheres w.r.t. the polar angle $\theta_{top}$ of the $t$~quark calculated from the decay products in the hadronic decay branch.
The direction measurement depends on the correct association of the $b$~quarks to the jets of the hadronic $W$~boson decays. 
The analysis is carried out separately for a left-handed polarised electron beam and for a right-handed polarised beam.
In case of a right-handed electron beam the direction of the $t$~quark can be precisely reconstructed. In case of a left-handed electron beam the final state features two hard jets from the $b$~quarks and soft jets from the hadronically decaying $W$~boson. This configuration leads to migrations in the polar angle distribution of the $\tpq$~quark as visible  in the left part of Fig~\ref{fig:ambig_rec}.
This implication motivates to restrict the determination of $\afbt$ in case of $\pem,\pep=-1,+1$ to cleanly reconstructed events. For this a test variable $\chi^2$ is defined that compares the measured values of the Lorentz factor $\gamma$ of the top, the momentum of the $b$~quark in the rest frame of the top and the angle $\mathrm{cos} \theta_{bW}$ between the $b$~quark and the $W$~boson.  The reconstructed polar angle distribution of the $t$~quark is compared with the generated one for different cuts on $\chi^2$. For a value of $\chi^2<15$ an excellent agreement between the generated and reconstructed polar angle distributions is obtained, see the right part of Fig.~\ref{fig:ambig_rec}. The tight selection however reduces the efficiency in case of left-handed initial electron beams from 55\% to 28\%. With this the forward backward asymmetry can be determined to a statistical precision of better than 2\%. The precise results corrected to the beam polarisations $\pem, \pep =\pm0.8,\mp0.3$ are given in Table~\ref{tab:resafb} together with those for the cross section, see previous section.  A more straightforward, albeit experimentally more challenging, way to control the migrations is to measure the charge of the $b$ quarks that are issue of the $t$ quark decay. References~\cite{amjad:tel-00949818} and~\cite{rouene:tel-01062136} describe the determination of the $b$~quark charge using secondary tracks.  The same value of $\afbt$ is obtained at a comparable selection efficiency~\cite{bib:these-rouene}. This means that $\afbt$ can be determined with two independent methods.  

\begin{table}[h!]
  \begin{center}
    \begin{tabular}{|c|c|c|}
      \hline
      $\pem, \pep$ & $(\delta \sigma / \sigma)_{stat.}$ [\%] & $(\delta \afbt / \afbt)_{stat.}$ [\%]\\ \hline
      $-0.8,+0.3$&  0.47 &1.8 \\ \hline
      $+0.8,-0.3$& 0.63 & 1.3\\ \hline
    \end{tabular}
  \end{center}
  \caption{\sl Statistical precisions expected for  the cross sections and $\afbt$ for different beam polarisations.}
  \label{tab:resafb}
\end{table}

Hard gluon radiation may alter the polar angle distribution of the final state $t$~quarks. The {\tt WHIZARD} version 1.95 used for the study generates hard gluons only via the interface to {\tt PYTHIA} that generates the parton shower. Therefore results presented before have been checked with a study on parton level using the most recent version 2.2.2 of {\tt WHIZARD} that correctly accounts for hard gluon radiation. No significant difference has been observed.



\section{Discussion of systematic uncertainties}

In the previous sections measurements of either cross sections or asymmetries have been presented. This section makes an attempt to identify and quantify systematic uncertainties,
which may influence the precision measurements. 

\begin{itemize}
\item \underline{Luminosity:} The luminosity is a critical parameter for cross section measurements only. The luminosity can be controlled to 0.1\%~\cite{Rimbault:2007zz}. 
\item \underline{Polarisation:} The polarisation is a critical parameter for all analyses. It enters directly the cross section measurements. The studies
presented in~\cite{bib:dbd:rosca} using $W$ pair production lead to an uncertainty of 0.1\% for the polarisation of the electron beam and to an uncertainty of 0.35\% for the polarisation of the positron beam. This translates into an uncertainty of 0.25\% on the cross section for $\pem,\pep=-0.8,+0.3$ and 0.18\% on the cross section for $\pem,\pep=+0.8,-0.3$. 
The uncertainty on the polarisation can be neglected with respect to the statistical uncertainty for $\afbt$.
\item \underline{Beamstrahlung and beam energy spread:} The mutual influence of the electromagnetic fields of the colliding bunches provokes radiation of photons known as {\em Beamstrahlung}. This Beamstrahlung modulates the luminosity spectrum, i.e.\,moves particles from the nominal energy to smaller energies. At the ILC for a centre-of-mass  energy of $500\GeV$ about 60\% of the particles are expected to have 99\% or more of the nominal energy~\cite{bib:ilc-tdr-dbd}. The beam energy spread, i.e.\,the RMS of this main luminosity peak is 124\,MeV for the electron beam and 70\,MeV for the positron beam~\cite{bib:ilc-tdr-dbd}. Both effects play a role at the $\ttbar$ threshold~\cite{Seidel:2013sqa} and can be neglected at energies well above this threshold.

\item \underline{Experimental uncertainties in top quark reconstruction:} As discussed in Sec.~\ref{sec:extrac} migrations have to be taken into account for the measurement of  $\afbt$, in particular for the
polarisations $\pem,\pep=-0.8,+0.3$. These migrations are reduced by stringent requirements on the event selection using a $\chi^2$ analysis. This in turn leads to a penalty in the efficiency. 
The success of the method depends in addition on a very sharp measurement of the variables used for the $\chi^2$ analysis. 
It is expected that these ambiguities can be (partially) eliminated by an event-by-event determination of the charge of the $b$~quark from the $t$~decay. As has been shown in Sec.~\ref{sec:exp}, the effect will be very much suppressed in case of $\pem,\pep=+0.8,-0.3$ beam polarisation.
\item \underline{Other experimental effects:} There is a number of other experimental effects such as acceptance, uncertainties of the $b$~tagging or the influence of passive detector material. The LEP1 experiments quote a systematic uncertainty on $R_b$ of 0.2\% a value which may serve as a guide line for values to be expected at the ILC, which on the other hand will benefit from far superior detector resolution and $b$~tagging capabilities. 
\item \underline{Theory:} 
The uncertainties of today's state-of-the-art calculations are discussed in
Section~\ref{sec:theory}. The uncertainties in the QCD corrections to
the total cross section and $A_{FB}$ are of the same order as the experimental 
uncertainties. Two-loop electro-weak calculations are required for a reliable estimation of the uncertainties due to electro-weak corrections. It is however intuitively clear that the latter will benefit from the 
insight of the different impact for different beam polarisations, see Sec.~\ref{sec:theory} and~\cite{Khiem:2015ofa}.  
\item \underline{Single-top production:} Single top production at the LC in 
association with a $W$~boson and bottom quark (through $WW^{\ast}$ production) 
leads to the same final state as $t$~quark pair production. Being largely suppressed in case of $\pem,\pep=+0.8,-0.3$ beam polarisation, it forms a 
sizeable contribution to the six-fermion final state in case of $\pem,\pep=-0.8,+0.3$ beam polarisation. It must therefore be taken 
into account in a realistic experimental strategy. This is left for a future study. 
\item \underline{Beyond Standard Model Physics:} Possible BSM effects may affect the various components of the background, in particular the $\ttbar$ induced background. This will therefore require a careful iterative procedure with tuning of event generators. This procedure seems feasible without a significant loss of accuracy.
\end{itemize}

As a summary it can be concluded that the total systematic uncertainties will not exceed the statistical uncertainties. This, however, requires an excellent control of a number of experimental and theoretical quantities. 



\section{Precision of form factors and electro-weak couplings}\label{sec:ff}

The measured cross sections and $\afbt$ lead  for two polarisation configurations to a set of four observables. By means of Eqs.~\ref{eq:sigma} and~\ref{eq:afbt} the uncertainties on these observables are used to build up a system of linear equations to determine the variances of up to four variables\footnote{For the Linear Algebra the software package {\tt Eigen}~\cite{bib:eigen-package} version 3.2.2 has been used.}, The variances are equivalent to the square of the standard deviations of the variables under study. These variables can be the form factors or alternatively directly the couplings. More explicitly, in this paper the following quantities will be determined separately:
\begin{enumerate}
\item The standard deviations of the Form Factors $\fonevA,\,\fonevZ,\,\foneaZ$ assuming no variation of the Form Factors $F^{X}_{2j}$;
\item The standard deviations of the Form Factors $\ftwovA,\,\ftwovZ$ assuming no variation of the Form Factors $F^{X}_{1j}$ ;
\item The standard deviations of the Couplings $\glA, \grA,  \glZ, \grZ$.
\end{enumerate}
Note, that the Form Factor $\foneaZ$ is fixed to be 0 in order to respect QED gauge invariance. On the other hand all four Couplings $g^{X}_{I} $are allowed to vary freely.
The resulting standard deviations are listed in Table~\ref{tab:results}. 

\begin{table}[h!]
\begin{footnotesize}
  \begin{center}
    \begin{tabular}{|c|c|c|c|c|c|c|c|c|c|}
      \hline
      Quantity & $\fonevA$ & $\fonevZ$  & $ \foneaZ$  & $\ftwovA$ & $\ftwovZ$ & $\glA$ &   $\grA$ & $\glZ$  & $\grZ$ \\ 
      \hline
      \hline
       SM Value at tree level & 2/3 &  0.230 & -0.595 & 0 & 0 & 2/3 & 2/3  & 0.824 & -0.364 \\ 
      \hline
       Standard deviation      & 0.002 &  0.003 & 0.007 & 0.001 & 0.002  & 0.005 & 0.005 & 0.008 & 0.009 \\ 
      \hline
      Relative precision [\%] & 0.3 &  0.9 & 1.2  & - & - & 0.8 & 0.8 & 1.0 & 2.5 \\ 
      \hline
    \end{tabular}
  \end{center}
  \caption{\sl Standard deviations and resulting relative precisions of form factors and couplings derived from the statistical precisions on the observables cross section and $\afbt$ as listed in Table~\ref{tab:resafb}.}
  \label{tab:results}
\end{footnotesize}
\end{table}
\noindent The complete covariance matrices are given in Appendix~\ref{app:cov}. From there it can be told that e.g. $\glZ$  and $\grZ$ are strongly correlated while $\fonevZ$  and $ \foneaZ$ are nearly uncorrelated.
\begin{figure}[tbp]
  \centering
\setlength{\unitlength}{1.5mm}

\begin{picture}(150,120)
\linethickness{0.3mm}
  \put(9,33){\line(0,1){2}}
  \put(5,34){\line(1,0){8}}  
  \put(15,34){...} 
  \put(21,34){\vector(1,0){60}} 
  \put(82,33){\Large{$\delta g^Z_R / g^Z_R$}}
  \multiput(31,33)(10,0){5}{\line(0,1){2}} 
  \put(6.5,31){\footnotesize{-330\%}} 
  \put(29,31){\footnotesize{-20\%}} 
  \put(39,31){\footnotesize{-10\%}}
  \put(60,31){\footnotesize{10\%}}
  \put(70,31){\footnotesize{20\%}}
  \put(51,4){\vector(0,1){60}} 
  \put(45,66){\Large{$\delta g^Z_L / g^Z_L$}}
  \multiput(50,14)(0,10){5}{\line(1,0){2}} 
  \put(45,13.5){\footnotesize{-20\%}} 
  \put(45,23.5){\footnotesize{-10\%}} 
  \put(45.5,43.5){\footnotesize{10\%}} 
  \put(45.5,53.5){\footnotesize{20\%}}
  \put(51,34){\color{red}\circle*{2}} 
  \put(53,36){\color{red}SM}
  \put(51,24){\color{britishracinggreen}\circle*{1.5}}
  \put(53,23){\color{britishracinggreen}Light top partners~\cite{Grojean:2013qca}}
  \put(41,24){\color{britishracinggreen}\circle*{1.5}} 
  \put(21,26){\color{britishracinggreen} Light top partners}
 \put(21,23){\color{britishracinggreen} Alternative 1~\cite{bib:panico-priv}}
  \put(76,59){\color{britishracinggreen}\circle*{1.5}} 
  \put(56,61.5){\color{britishracinggreen} Light top partners Alternative 2~\cite{bib:panico-priv}}
   \put(51,19){\color{cyan}\circle*{1.5}} 
   \put(53,18){\color{cyan}Little Higgs~\cite{Berger:2005ht}}
  \put(51,14){\color{gray}\circle*{1.5}} 
  \put(53,13){\color{gray}RS with Custodial SU(2)~\cite{Carena:2006bn}}
  \put(51,9){\color{orange}\circle*{1.5}} 
  \put(53,8){\color{orange}Composite Top~\cite{Pomarol:2008bh}}
  \put(31,14){\color{magenta}\circle*{1.5}} 
  \put(20,16){\color{magenta}5D Emergent~\cite{Cui:2010ds}}
  \put(56,29){\color{camel}\circle*{1.5}} 
  \put(58,28){\color{camel}4D Composite Higgs Models~\cite{Barducci:2015aoa}}
  \put(9,34){\color{blue}\circle*{1.5}} 
  \put(1,36){\color{blue}RS with Z-Z' Mixing~\cite{Djouadi:2006rk}}
\multiput(56,40)(30,0){2}{\line(0,1){12}} 
\multiput(56,40)(0,12){2}{\line(1,0){30}}
\put(61,47){\large{ILC Precision}}
\color{red}
\qbezier(73.3463, 43.7167)(73.2699, 43.9671)(72.5286, 43.9342)
\qbezier(72.5286, 43.9342)(71.7873, 43.9013)(70.8154, 43.6044)
\qbezier(70.8154, 43.6044)(69.8435, 43.3075)(69.2103, 42.9205)
\qbezier(69.2103, 42.9205)(68.5772, 42.5336)(68.6537, 42.2833)
\qbezier(68.6537, 42.2833)(68.7301, 42.0329)(69.4714, 42.0658)
\qbezier(69.4714, 42.0658)(70.2127, 42.0987)(71.1846, 42.3956)
\qbezier(71.1846, 42.3956)(72.1565, 42.6925)(72.7897, 43.0795)
\qbezier(72.7897, 43.0795)(73.4228, 43.4664)(73.3463, 43.7167)
\end{picture}

\caption{\sl Predictions of several models that incorporate Randall-Sundrum (RS) models and/or compositeness or Little Higgs models on the deviations of the left- and right-handed couplings of the $t$~quark to the $Z^0$ boson. The ellipse in the frame in the upper right corner indicates the precision that can be expected for the ILC running at a centre-of-mass energy of $\roots = 500\GeV$ after having accumulated ${\mathcal L}=500\,\invfb$ of integrated luminosity shared equally between the beam polarisations $\pem,\,\pep =\pm0.8,\mp0.3$. The original version of this figure can be found in~\cite{Richard:2014upa}.}
\label{fig:models-rp}
\end{figure}
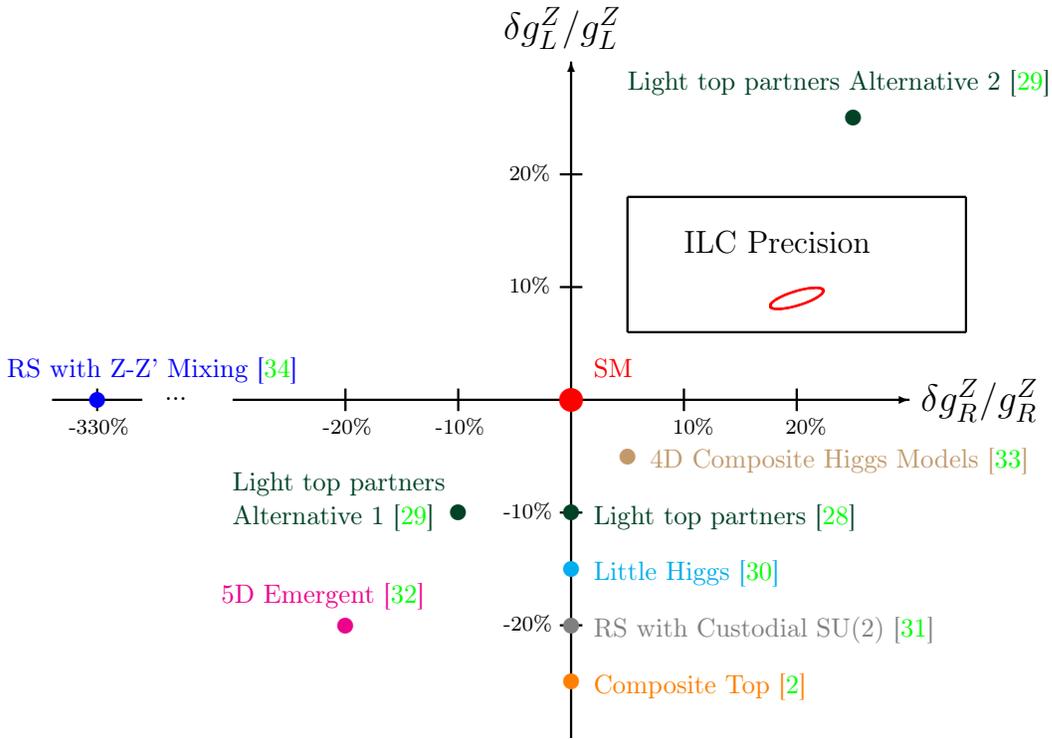
The expected high precision at a linear $\epem$ collider allows for a profound discussion of effects by new physics. The findings can be confronted with predictions in
 the framework of Randall-Sundrum models and/or compositeness models such as~\cite{Pomarol:2008bh,Djouadi:2006rk,Hosotani:2005nz,Cui:2010ds,Carena:2006bn,Grojean:2013qca,bib:panico-priv,Barducci:2015aoa} or Little Higgs models as e.g.~\cite{Berger:2005ht}. All these models entail deviations from the Standard Model values of the $t$~quark couplings to the $Z^0$ boson that will be measurable at the ILC as illustrated in Fig.~\ref{fig:models-rp}. Therefore, the couplings of the  $t$~quark to the $Z^0$ boson will discussed in a broader context in the following.



\subsection{Discussion of couplings to the $Z^0$ Boson - Comparison with perspectives for LHC and Flavour Physics}
\begin{figure}
\centering
\includegraphics[width=0.5\columnwidth]{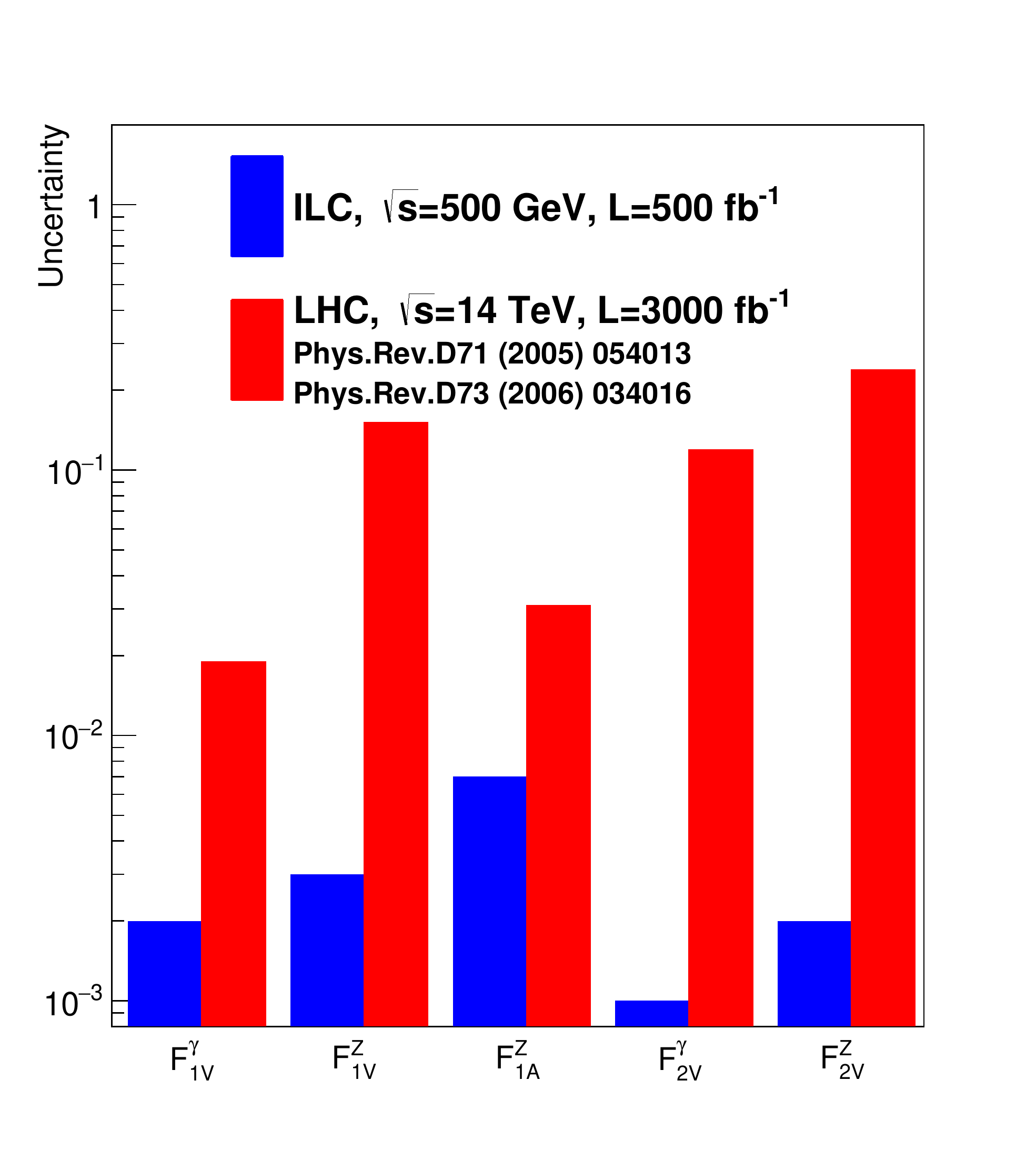}
\caption{\sl
Graphical comparison of statistical precisions on $CP$ conserving form factors
 expected at the LHC, taken from~\cite{Baur:2004uw}  and~\cite{Baur:2005wi}, and at the ILC. 
The LHC results assume an integrated luminosity of $\mathcal{L}=3000~\invfb$ at 
$\roots=14\,\TeV$.
The results for the ILC assume an integrated luminosity of $\mathcal{L}=500~\invfb$ at 
$\roots=500\GeV$ and a beam polarisation  $\pem=\pm0.8,\pep = \mp0.3$.}
\label{fig:couplilclhc}
\end{figure}

Electro-weak couplings can be measured at the LHC in associated $\bar{t}t \gamma$ and $\bar{t}tZ$ production. A comprehensive compilation on the statistical precisions on the form factors that can be expected at the end of the HL-LHC is given in~\cite{Baur:2004uw} and~\cite{Baur:2005wi} for an update on $\bar{t}tZ^0$ form factors. The values published there are compared with the results in the present study in Fig.~\ref{fig:couplilclhc}.  
All but one form factor will be measured at about a factor 10 better at the ILC for the scenario discussed in this paper than it will be possible at the LHC. This exception is $ \foneaZ$ where~\cite{Baur:2005wi} quotes a possible statistical precision of $\delta \foneaZ \approx 0.031$.  It should however be pointed out that the considerable precision expected for  $\delta \foneaZ$ benefits strongly from LEP/SLC bounds on the oblique parameters that e.g. render it unlikely that $ \foneaZ$ flips sign due to New Physics.
The study presented by~\cite{Baur:2005wi} is an analysis at leading order QCD. The analysis carried out in~\cite{Rontsch:2015una} suggests that higher-order effects in the theory may allow for an improvement of the LHC precision by up to 40\%. Note at this point that the interference between the $\gamma$ and the $Z^0$ in case of $e^+e^- \rightarrow t\bar{t}$ will allow for measuring flips of the signs of the form factors that will be unnoticed in associated $\bar{t}tZ^0$ at the LHC.

While the prospects for the LHC discussed so far are based on analyses differential in given jet observables of the final state, LHC experiments observe the process $pp \rightarrow \bar{t} t Z^0$~\cite{Chatrchyan:2013qca,Khachatryan:2014ewa,collaboration:2014zl,bib:cms-top-14021}. The interpretation of the results is however still limited by the small statistics available for the analyses.  
At the LHC electro-weak couplings are measured also in single $t$~quark production. In the effective field theory approach, assuming $SU(2)_L \times U(1)_Y$ gauge symmetry for the operators, the relation 
\beq
\frac{\delta  g^{tbW}_L} {g^{tbW}_L}  \approx  0.35 \frac{\delta \glZ} {\glZ }
\label{eq:relnccc}
\eeqn 
can be established. Here $g^{tbW}_L$ is the charged current coupling of the decay $t \rightarrow W b$.
The CMS Collaboration~\cite{Khachatryan:2014iya} reports a precision for the $t$-$b$ transition probability $V_{tb}$ of about 4\%. In the Standard Model $V_{tb}$ is identical to $g^{tbW}_L$. Hence, by means of Eq.~\ref{eq:relnccc} the precision of the coupling of left-handed $t$~quarks to the $Z$ boson can be derived to be of the order of 11\%.
Noting that $\sigma(pp \rightarrow \bar{t} t Z) \sim (g^{Z}_L)^2 +  (g^{Z}_R)^2$ this allows in principle also for deriving  $(\grZ)^2$, albeit with a poor precision given that $(\glZ)^2 \gg (\grZ)^2$. 
Loop corrections in heavy flavour physics as e.g. in the processes $b \rightarrow s \gamma$, $B\rightarrow \mu^+ \mu^-$ or $K\rightarrow \mu^+ \mu^-$, respectively, may also lead to competitive determinations of $\delta \glZ$~\cite{Brod:2014hsa}. However, again $\grZ$ can only be constrained rather poorly.

It follows that the ILC will allow for measurements superior to those that can be expected from existing experiments. This is particularly true for the determination of    $\delta \grZ$. 

\subsection{Coupling measurements and form factors at different centre-of-mass energies}

Given the fact that at the ILC in its current layout centre-of-mass energies of up to $1\TeV$ can be reached and that the alternative project for a linear collider, CLIC~\cite{Lebrun:2012hj}, may even reach higher energies, it is instructive to discuss the results presented in this paper with this possibility in mind.  The selection
and reconstruction of the decay topology of boosted $t$~quarks is very different from that of
$t$~quarks with moderate velocity. Therefore, the study must be extrapolated to high centre-of-mass energy with some care. 
Still, the following observations can be made:
\begin{itemize}
\item Neglecting varying detector systematics and theory uncertainties with varying centre-of-mass energy, and assuming the linear
collider luminosity vs. centre-of-mass energy curve, the sensitivity to the form factors considered
in this paper is greatest at approximately $\roots=400-700\GeV$. At lower centre-of-mass
energy, as e.g. studied in~\cite{Janot:2015yza}, the small velocity of the $t$~quarks reduces the potential of the $\afbt$ measurement compromising thus the measurement of the 
axial vector couplings to the $Z^0$~boson and by virtue of Eq.~\ref{eq:qtrl} the disentangling of left- and right-handed couplings.
On the other hand running at centre-of-mass energies close to the $\ttbar$ threshold offers sensitivity to virtual Higgs exchange~\cite{Harlander:1995dp,Atwood:2000tu,Horiguchi:2013wra}. In case the Higgs has a $CP$ odd component  this may give rise to recognisable $CP$ violating effects in the threshold region~\cite{Bernreuther:1996jk}.
However, in the transition region between the $\ttbar$ threshold and the continuum region starting at around $380\GeV$ the current QCD uncertainties are at least 10\%. This is due to uncertainties on higher QCD order corrections and on the correct matching procedure between the non-relativistic calculations at the $\ttbar$ threshold and the relativistic continuum calculations~\cite{bib:privbhs}. 
\item If an effect is seen at  $\roots=500\GeV$ it will be crucial to know how it evolves with energy with a decent lever arm. If, for instance, the effect is due to mixing of the $Z^0$ boson with a new $Z'$ boson it will remain unchanged. If, however, a $Z'$ boson leads to a propagator term, the corresponding effect will grow like $s/M^2_{Z'}$. In the case of Randall-Sundrum Models both effects are present and therefore measurements at two energies are needed to extract $M_{Z'}$, see e.g.~\cite{Richard:2014upa} for a deeper discussion.
\item The impact of high-scale new physics on the observables can increase strongly with
centre-of-mass energy. Operators corresponding to the top quark dipole moments
and four-fermion contact interactions induce larger anomalous form factors at higher energy.
For other anomalous couplings, however, the impact is nearly independent
of the centre-of-mass energy as is the case for $\fonevZ$ and $\foneaZ$.
\end{itemize}
A full simulation study at different centre-of-mass energies is left for a future publication.
\section{Summary and outlook}

This article presents a comprehensive analysis fully simulated events of $\ttbar$~quark production at the International Linear Collider using the semi-leptonic decay channel. Results are given for a centre-of-mass energy of $\roots = 500\GeV$ and an integrated luminosity of $\mathcal{L}=500\,\invfb$ shared equally between the beam polarisations $\pem,\,\pep=\pm0.8,\mp0.3$.  

Semi-leptonic events, including those with $\tau$ leptons in the final state can be selected with an efficiency of about 55\%.
The cross section of the semi-leptonic channel of $\ttbar$~quark production can therefore be measured to a statistical precision of about 0.5\%. The second observable is the forward-backward asymmetry $\afbt$. It was shown that in particular for predominantly left-handed polarisation of the initial electron beam the $V-A$ structure leads to migrations, which distort the theoretical expected $\afbt$. These migrations can be remedied by tightening the selection criteria of the events or alternatively by measuring the charge of the $b$ quark produced in the decay of the $t$ quark. Taking into account this correction the forward-backward asymmetry can be determined to a statistical precision of better than 2\% for both beam polarisations. 

The observables together with the unique feature of the ILC to provide polarised beams allow for a largely unbiased disentangling of the individual couplings of the $t$~quark to the $Z^0$~boson and the photon. These couplings can be measured with high precision at the ILC and, when referring to the results in~\cite{Baur:2004uw,Baur:2005wi}, considerably better than it will be possible at the LHC even with an integrated luminosity of $\mathcal{L}=3000\,\invfb$. The improving analyses of the LHC experiments will however be observed with great interest.

Beam polarisation is a critical asset for the high precision measurements of the electroweak $t$ quark couplings. Experimental and theoretical effects  manifest themselves differently
for different beam polarisations. It seems to be that the configuration $\pem,\,\pep=+0.8,-0.3$ is more benign in both, experimental aspects due to the suppression of migrations in the polar angle spectrum of the final state $t$ quark and theoretical aspects due to the somewhat simpler structure of higher order electroweak corrections.
It is intuitively clear that the described facts would greatly support the discovery of effects due to new physics.

The precision as obtained in the present study for the ILC would allow for the verification of a great number of models for physics beyond the Standard Model. Examples for these models are extra dimensions and compositeness. The results obtained here constitute therefore a perfect basis for discussions with theoretical groups. Note at this point that running scenarios for the ILC have been proposed that would yield between 8 and 10 times more integrated luminosity~\cite{Barklow:2015tja} than it is assumed for the present study.  Moreover it can be expected that the event reconstruction will be improved by e.g.\,the measurement of the $b$~quark charge. It is therefore not statistics that will limit the final accuracy but most likely theory and experimental systematics.

Hence, the study of systematic errors, only partially addressed in this study, will become very important. Already from the achieved precision it is mandatory that systematics are controlled to the 1\% level or better in particular for the measurement of the cross section. This issue is addressed in ongoing studies. 

The study presented in~\cite{Khiem:2015ofa} based on generated events suggests that by exploiting the polarisation of the final state $t$ quarks a simultaneous extraction of all ten form factors, see Eq.~\ref{eq:vtxvtt}, to a precision below the percent level is feasible. A detailed comparison between the advantages and drawbacks of the method applied there and the method presented in this paper is left for a future study.

\section*{Acknowledgements}

This work was supported within the 'Quarks and Leptons' programme of the CNRS/IN2P3, France, and by the Spanish Grant Agreement "Detector developments and physics
studies for future colliders, FPA2012-39055-C02-01".
The results benefit from the enlightening discussions in the framework of the French-Japanese  FJPPL/TYL 'virtual laboratory'  on top physics, particularly through comments by Emi Kou, Keisuke Fujii and Fran{\c c}ois LeDiberder. We would like to thank Fabian Bach and Maximilan Stahlhofen for the profound discussion on QCD uncertainties in the $\ttbar$ threshold region.

\begin{appendices}
\section{Covariance matrices} \label{app:cov}
For completeness the underlying covariance matrices of the results presented in Sec.~\ref{sec:ff} are given in this appendix.

\begin{itemize}
\item The covariance matrix resulting from the system of linear equations built for the Form Factors $\fonevA,\,\fonevZ,\,\foneaZ$ reads:

\beq
 \left[ {\begin{array}{ccc}
  var(\fonevA) &   cov(\fonevA,\fonevZ)  & cov(\fonevA,\foneaZ) \\
                         &   var(\fonevZ)                 & cov(\fonevZ,\foneaZ) \\
                         &                                          & var(\foneaZ)
  \end{array} } \right]
=
   \left[ {\begin{array}{ccc}
   0.260  &   -0.043  & 0.506 \\
                         &   0.791   & 0.118 \\
                         &                 & 5.460
  \end{array} } \right] \times 10^{-5}.
\label{eq:covf1}
\eeqn

\item The covariance matrix resulting from the system of linear equations built for the Form Factors $\ftwovA,\,\ftwovZ$ reads:
\beq
 \left[ {\begin{array}{cc}
  var(\ftwovA) &   cov(\ftwovA,\ftwovZ)   \\
                         &   var(\ftwovZ)                  \\
  \end{array} } \right]
=
   \left[ {\begin{array}{ccc}
   0.160  &   -0.046   \\
                 &   0.473   \\

  \end{array} } \right] \times 10^{-5}.
\label{eq:covf2}
\eeqn

\item The covariance matrix resulting from the system of linear equations built for the Couplings $\glA, \grA,  \glZ, \grZ$ reads:

\begin{eqnarray}
\nonumber
 \left[ {\begin{array}{cccc}
  var(\glA) &   cov(\glA,\grA)  & cov(\glA,\glZ) & cov(\glA,\grZ) \\
                         &   var(\grA)     & cov(\grA,\glZ) &  cov(\grA,\grZ)\\
                         &                        & var(\glZ)         & cov(\glZ,\grZ) \\
                         &                        &                         & var(\grZ) 
  \end{array} } \right]
=\\  
   \left[ {\begin{array}{cccc}
   2.801&   -2.357  & -0.339 & -1.234 \\
              &   2.949   & -0.780 &  2.247 \\
              &                 & 5.988   & -4.691\\
              &                 &               & 7.306 
  \end{array} } \right] \times 10^{-5}.
\label{eq:covg}
\end{eqnarray}

\end{itemize}


\end{appendices}
\section*{References}
\bibliographystyle{utphys_mod}
\begin{footnotesize}
\bibliography{tt-pap}
\end{footnotesize}

\end{document}